\documentclass[aps,prl,showpacs,floatfix,twocolumn,superscriptaddress]{revtex4-1}
\usepackage{graphicx}
\usepackage{bm,amsmath}
\usepackage{dcolumn}
\usepackage{amsfonts,amssymb}

\newcommand{\beq}{\begin{eqnarray}}
\newcommand{\eeq}{\end{eqnarray}}

\newcommand{\avg}[1]{\left\langle#1\right\rangle}

\newcommand{\ket}[1]{\left| #1\right\rangle}

 \newcommand{\be}{\begin{equation}}
 \newcommand{\ee}{\end{equation}}

\begin{document}

\title{Confinement in the bulk, deconfinement on the wall: \\ infrared equivalence between compactified QCD and quantum magnets}

\author{Tin Sulejmanpasic} 
\email{tin.sulejmanpasic@gmail.com}
\affiliation{Philippe Meyer Institute, Physics Department, \'Ecole Normale Sup\'erieure, PSL Research University, 24 rue Lhomond, F-75231 Paris Cedex 05, France}
\affiliation{Department of Physics, North Carolina State University, Raleigh, North Carolina 27695, USA}
\author{Hui Shao}
\email{shaohui@csrc.ac.cn}
\affiliation{Beijing Computational Science Research Center, Beijing 100193, China}
\affiliation{Department of Physics, Boston University, Boston, Massachusetts 02215, USA}
\author{Anders W. Sandvik}
\affiliation{Department of Physics, Boston University, Boston, Massachusetts 02215, USA}
\author{Mithat \"Unsal} 
\affiliation{Department of Physics, North Carolina State University, Raleigh, North Carolina 27695, USA}
\date{\today}

\begin{abstract}
In a spontaneously dimerized quantum antiferromagnet, spin-$1/2$ excitations (spinons) are confined in pairs by strings akin to
those confining quarks in non-abelian gauge theories. The system has multiple degenerate ground states (vacua) and domain walls
between regions of different vacua. For two vacua, we demonstrate that spinons on a domain wall are liberated, in a
mechanism strikingly similar to domain-wall deconfinement of quarks in variants of quantum chromodynamics. This observation
not only establishes a novel phenomenon in quantum magnetism, but also provides a new direct link between
particle physics and condensed-matter physics. The analogy opens doors to improving our understanding of particle confinement
and deconfinement by computational and experimental studies in quantum magnetism.
\end{abstract}
\maketitle

The phenomenon of confinement is well known in quantum chromodynamics (QCD), where quarks are bound by 'strings' and can only
be observed within composites; the mesons and baryons. The physics underlying confinement is still poorly understood, e.g.,
as concerns the nature of the confining strings, because the relevant 3+1 dimensional (D) non-abelian gauge theories are strongly
coupled and reliable analytical methods are lacking. Numerical lattice calculations with strings are also challenging, especially
in the presence of matter.

Some understanding of confinement has been developed within supersymmetric (SUSY) gauge theories, which generically have multiple vacua. Important for this work is a conjecture due to Rey and advocated by
Witten \cite{Witten:1997ep} that confining SUSY gauge theories facilitate deconfinement of quarks on domain walls interpolating between two vacua. Recently \cite{Anber:2015kea} by utilization of the special kind of compactification \cite{Unsal:2007jx,Unsal:2008ch} it was demonstrated explicitly that this feature transcends SUSY theories and is generic for QCD-like theories. In this work we show that the \emph{liberation on the walls} also transcends QCD-like theories and takes place in quantum magnets, a setting that may very well be experimentally realizable.

In condensed matter physics, certain excitations can be regarded as composites of confined objects, and in some cases deconfinement,
or fractionalization, takes place. The most famous example is that of charge $e/3$ excitations in the fractional quantum Hall
effect \cite{tsui82,laughlin83}. Another well-established case is the liberation of spinons carrying spin $S=1/2$ in
antiferromagnetic spin chains \cite{faddeev81,shastry81,tennant93}, in contrast to the $S=1$ spin waves in higher dimensions.
Here we focus on a system with close correspondence in gauge theories: a spontaneously dimerized 2D quantum magnet (a valence-bond-solid,
VBS), where spins paired up into localized singlets form a crystalline pattern, e.g., in columns on the square lattice \cite{read89}.
Due to lattice symmetries, the pattern can form in multiple ways, corresponding to different vacua. When exciting such a state by
breaking a bond, the two unpaired (or triplet-paired) spins are confined by a string of deformed VBS texture. Deconfinement can
take place if the VBS is weakened upon approaching a so-called deconfined quantum-critical point \cite{senthil04a,senthil04b}. However,
the identity of the spinon as a quasi-particle is lost at the critical point, due to the gapless critical host system \cite{tang13}.
Truly deconfined spinons are believed to exist in gapped, topological spin liquid phases \cite{balents10}.

\begin{figure}[tp]
\centering
\includegraphics[width=\columnwidth]{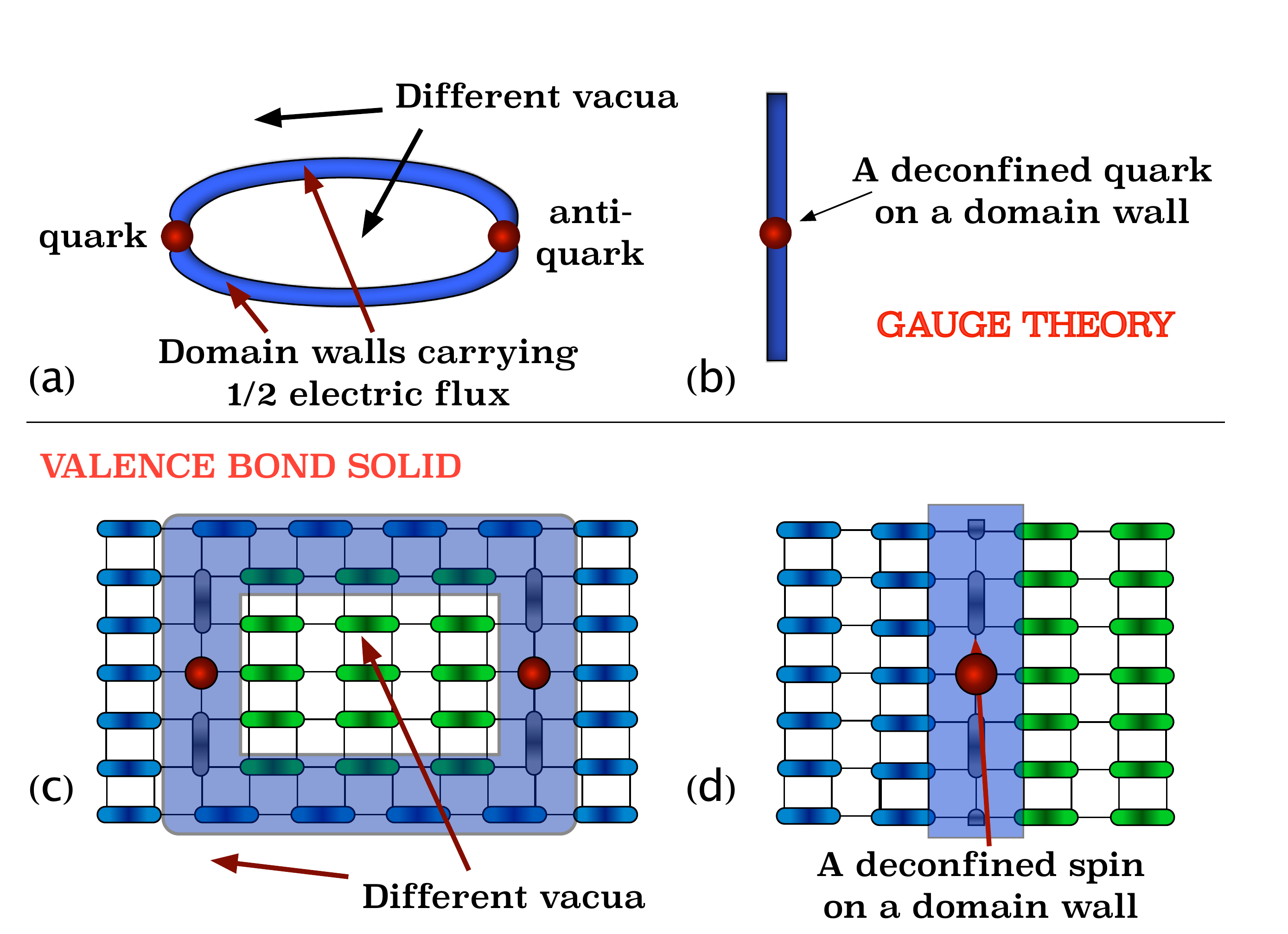}
\caption{Deconfinement on domain walls. A quark-confining string (a) composed out of two strands (domain walls) separating
two vacua \cite{Anber:2015kea} allow deconfinement on a domain wall (b). Confinement in a VBS with $Z_2$ degeneracy, where two
domain walls form between unpaired spins (c). A domain wall absorbs the string completely, liberating the spinons (d).}
\label{fig:deconf}
\end{figure}

Here we point out that the analogy between quark and spinon confinement is not superficial, but the two phenomena can be described
in strikingly similar terms, as illustrated in Fig.~\ref{fig:deconf}. We describe a mechanism of spinon liberation on a VBS domain wall
and demonstrate this explicitly by quantum Monte Carlo (QMC) simulations of a spin model hosting $Z_4$ (four vacua) or $Z_2$ (two vacua)
VBS ground states. Deconfinement on the domain wall takes place upon reducing the $Z_4$ symmetry to $Z_2$. Given the highly non-perturbative
nature of confinement, the ability to study this phenomenon in the setting of quantum magnets, in simulations and potentially also in experiments,
is a promising avenue for making further progress.

The previous work on quark deconfinement on a domain wall \cite{Anber:2015kea} required a new (non-thermal) compactification of the gauge theory,
$\mathbb{R}^{3,1}\rightarrow \mathbb{R}^{2,1} \times S^1$, where $S^1$ is a spatial direction (i.e., the long distance theory is 2+1D) \cite{Unsal:2007jx,Unsal:2008ch}.
The liberation phenomenon relied on the fact that the defects causing confinement have magnetic charge $\pm 2 $ \cite{Note0}, leading to stable line-like domain walls that constitute the confining strings and carry a half unit of electric flux. Since quarks carry a whole unit of electric charge,
quark--anti-quark pairs are bound by two separate half-flux strings. The fact that fundamental strings can end on domain walls is then a matter of
geometry: strings confining the electric charges consist of two domain walls carrying half the electric flux each. The string can then expand
and form a domain wall, as illustrated in the top panel of Fig \ref{fig:deconf}.

All of these properties are manifested within the following Euclidean Lagrangian
\cite{Unsal:2007jx,Unsal:2012zj,Anber:2015kea}:
\begin{equation}
\label{eq:YMeff}
 L_{\rm eff}=M \left[(\partial_\mu\chi)^2-m^2{\cos(2\chi)}\right]\;,
\end{equation}
where $\chi+2\pi= \chi$ is an angular field referred to as the \emph{dual photon}, $M\sim g(L)/L$ and $m^2\sim 1/L^2e^{-8\pi^2/g^2(L)}$, with $L$ the size of
the compact direction and $g(L)$ the running gauge coupling at scale $L$. The dual photon arises upon compactification from $3+1D$ to $2+1D$ because the
gauge group abelianizes to U(1) [starting from an SU(2) gauge group]. Normally one would expect a term $e^{i\chi}+e^{-i\chi}$ in $L_{\rm eff}$, but
if there are massless adjoint fermions these contributions are forbidden \cite{Unsal:2007jx} due to the Nye-Singer index theorem \cite{Nye:2000eg}, and the first term to contribute is the so-called \emph{the magnetic bion} $e^{\pm 2i\chi}$ \cite{Unsal:2007jx}.
Alternatively, Eq.~\eqref{eq:YMeff} arises from a pure Yang-Mills theory with topological $\theta$-angle term. In that case there are two distinct monopole and
anti-monopole events, with magnetic and topological charges $(Q_m,Q_{top})$ $=$ $(1,1/2)$, $(-1,1/2)$, $(1,-1/2)$, $(-1,-1/2)$. Their couplings
$e^{i\chi+i\frac{\theta}{2}}$, $e^{-i\chi+i\frac{\theta}{2}}$, $e^{-i\chi-i\frac{\theta}{2}}$, $e^{i\chi-i\frac{\theta}{2}}$ to the $\chi$-field and
the $\theta$-angle add up to $\cos(\theta/2)\cos(\chi)$. By setting $\theta=\pi$, this term vanishes via topological interference
\cite{Unsal:2012zj}, and the magnetic bion mechanism again yields \eqref{eq:YMeff}. The reader is referred to the literature for
details, e.g., the review \cite{Dunne:2016nmc}.  In quantum magnets, the crucial role of index theorem and topological interference, which leads to \eqref{eq:YMeff},   is played by Berry phase induced interference (see below).

In the path integral, wordlines of charged particles (i.e. quarks) act like $2\pi$-vortices in the field $\chi$, i.e., the winding of the compact field $\chi$ around
the quark measures its \emph{chromo-electric charge}. The potential $-\cos(2\chi)$, however, forces the $\chi$-field to settle at either $\chi=0$
or $\chi=\pi$; the two vacua of the SU(2) theory \cite{Note1}. A quark is then attached to two collimated $1/2$ flux strings, across which the
$\chi$-field winds by $\pi$; see Fig. \ref{fig:deconf}(b). These half-flux strings are domain walls, which has a remarkable, but simple
consequence that the insertion of a confined quark generates a domain wall containing that quark. Consequently, the quark can move freely along
the domain wall.

\begin{figure}[tp]
   \centering
   \vspace{0.4cm}
   \includegraphics[width=0.8\columnwidth]{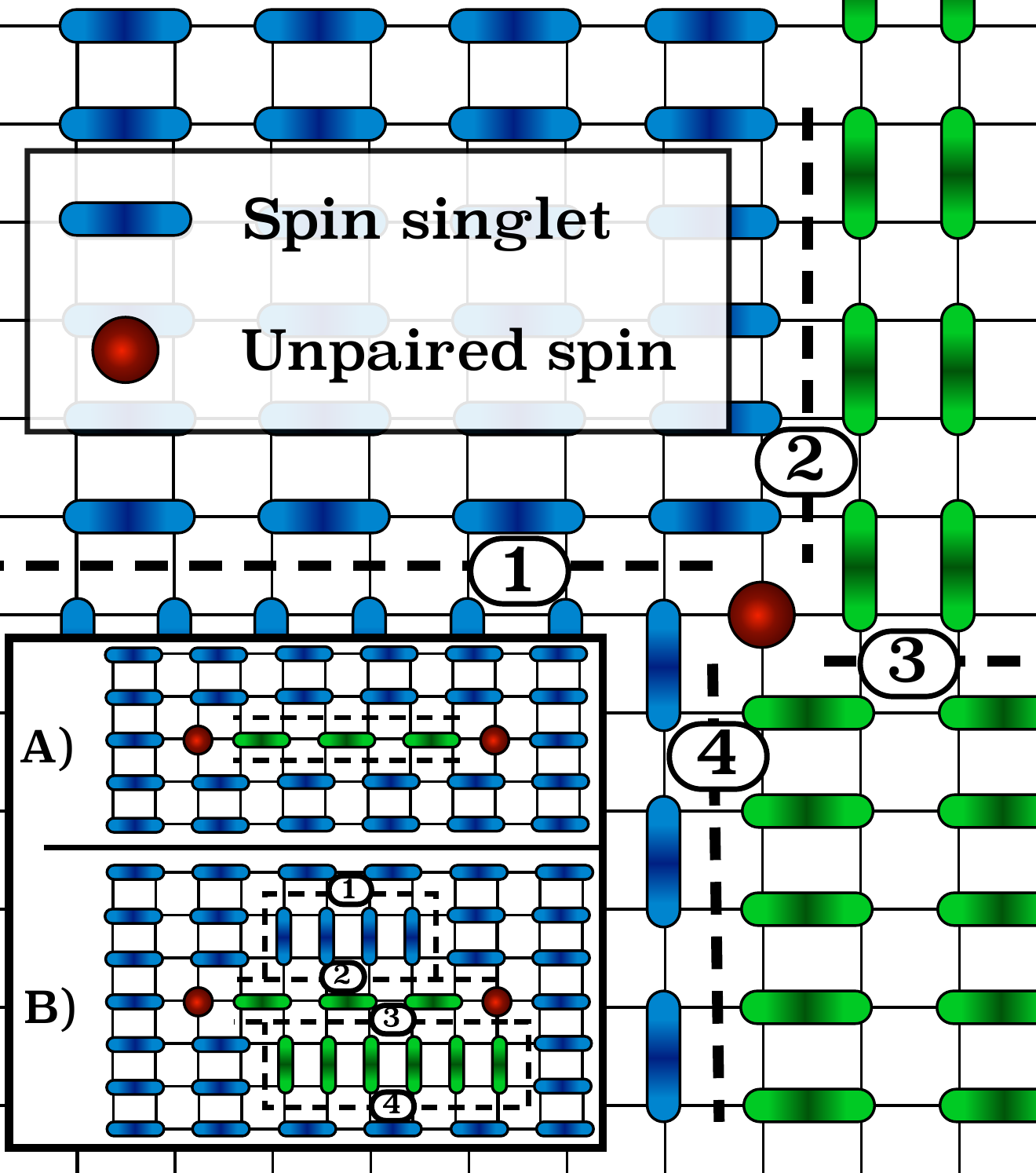}
   \caption{The vacua of a $Z_4$ VBS, illustrated with short valence bonds (singlets). Petterns
with a relative shift of one lattice spacing are colored with blue and green.  The four phases meet at an unpaired spin (red circle).
A spinon can be thought of as such a nexus of four different kinds of domain walls (dashed lines labeled by the numbers 1-4) with a spin
in the core. Spinons are then not confined by a single string (inlay A), but by \emph{four} string-like domain lines (inlay B).}
   \label{fig:main}
\end{figure}

We now discuss an analogous phenomenon in a VBS quantum magnet. The columnar VBS on the uniform square lattice breaks $Z_4$ symmetry,
leading to domain walls when boundary conditions force different patterns (vacua) in different parts of the system. The four vacua (which can be associated
with $\chi=0,\pi/2,\pi,3\pi/2$) can meet at a single point, in which case the presence of an unpaired spin (spinon) is required at the nexus \cite{levin04},
as illustrated in Fig.~\ref{fig:main}. A domain wall between, say, the two different horizontal dimer patterns, representing a $\pi$ winding of $\chi$,
will split into two $\pi/2$ domain walls separated by a region with vertical dimers \cite{shao15}.

Confinement of spinons in the VBS is now a matter of topology: an unpaired spin causes misalignment of dimers, forcing interfaces between
inequivalent vacua. Two unpaired spins must be connected by two defect lines, which are domain walls separating the two vacua
(inlay A of Fig. \ref{fig:main}). However, since four domain walls intersect at an unpaired spin, two unpaired spins are also connected
by four domain walls. The picture in inlay B of Fig.~\ref{fig:main} is therefore a more accurate illustration of the composite nature
of strings. We should emphasize however, that the spinons are necessarily dynamical, and the string will break once its energy content becomes
comparable to the mass of the $S=1$ excitation, so that new pairs can be created (in analogy to meson and baryon creation upon separating quarks).
This is in contrast to the gauge theories we discussed and to the quantum dimer model, where there is no internal spin structure of the dimers and
the strings are therefore stable and indeed show domain wall structure \cite{Banerjee:2015pnt}. The dimer model can be thought of as a pure gauge
theory without matter fields, while the full quantum magnet inseparably contains matter.

In the gauge theory description of the VBS \cite{haldane1988,read89,murthy90,senthil04a,senthil04b}, spins are represented by vectors on the
Bloch sphere coupling to Berry phases. An antiferromagnet is described by a unit vector field $\hat n$ on the spatial lattice in continuous time.
Haldane \cite{haldane1988} showed that the Berry phase in 2+1D has no influence on smooth $\hat n$ configurations. However, it couples to singular
``hedgehog'' configurations in space-time. These hedgehogs can disorder the system, rendering $\avg{\hat n}=0$, e g., in the VBS phase
\cite{read89,murthy90}. The hedgehogs are the analogues of the monopole-instantons, which had a profound influence on the gauge dynamics discussed
above. On the square lattice these events appear on the dual lattice (i.e., centers of plaquettes) and couple to Berry phases as
$1,e^{i\frac{\pi}{2}},e^{i{\pi}},e^{i{3\pi/2}}$, depending on which of the four sublattices of the dual lattice they occupy \cite{haldane1988}.
Further one can write $\hat n(x)=u^\dagger(x) \vec\sigma u(x)$, where $x$ is a position on the lattice and  $u(x)=(u_1(x),u_2(x))$ is a bosonic or
fermionic SU(2) doublet, with the constraint ${u_1}^\dagger u_1+{u_2}^\dagger u_2=1$. This parametrization is invariant under the local gauge
rotation $u\rightarrow e^{i\alpha} u$, and the effective theory with the operators $u_{1,2}$ is therefore a U(1) gauge theory.
In the path-integral the hedgehog configurations of $\hat n$ appear as monopoles of this U(1) gauge group.

N\'eel order implies that $u$ ``condenses'', "breaking" the U(1) gauge symmetry spontaneously. In the absence of N\'eel order, $u$ can be
integrated out and the remaining pure gauge theory can be dualized to a single compact scalar field $\chi$, as before. In this case there are
four types of monopoles (and their anti-monopoloes), coupling to the $\chi$ field and Berry phases as $e^{i\chi+i\frac{l\pi}{2}}$ and $e^{-i\chi-i\frac{k\pi}{2}}$
($k=0,1,2,3$). However, only 4-monopole events are possible \cite{read89,murthy90}, by reasons very similar to those discussed around Eq.~(\ref{eq:YMeff}).
The potential $\cos(4\chi)$ then forms and leads to four distinct vacua labeled by $\chi=0,\pi/2,\pi,3\pi/2$. Domain walls interpolate between vacua
with $\chi$ and $\chi+\pi/2$ and carry energy proportional to their length. The insertion of an unpaired spin at $x$ amounts to inserting $u_1^\dagger(x)$
(spin up) or $u_2^\dagger(x)$ (spin down). For definiteness, let $u_{1,2}$ be fermionic, and label the set of states by the occupation
numbers $u_1^\dagger u_1$ and $u^\dagger_2 u_2$ as $\Omega=\{\ket{00},\ket{01},\ket{10},\ket{11}\}$. The constraint $u_1^\dagger u_1+u_2^\dagger u_2=1$
is obeyed only by the states $\ket{10}$ (up) and $\ket{01}$ (down) at the given site. Insertion of $u^\dagger_{1,2}(x)$ therefore ensures that the
spin at $x$ is up or down, because upon projection to physical spin states $u^\dagger_1\Omega,u^\dagger_2\Omega$ contain only $\ket{10},\ket{01}$.
Since $u_{1,2}^\dagger(x)$ are charged under the U(1) gauge group, they represent the fictitious electric charges, which impose winding by $2\pi$ on
the $\chi$ field. An isolated unpaired spin then sources the four domain walls as in Fig.~\ref{fig:main}.

The $Z_4$ VBS phase does not allow for spinons to be deconfined on the domain wall, as a single domain wall, say between $\chi=0$ to
$\chi=\pi/2$, would absorb only two out of four domain walls to which an isolated spin is attached. The remaining two cause confinement.
If the Lagrangian is deformed, however, to change the ground state degeneracy from 4 to 2, e.g., with horizontal singlets energetically
preferred in Fig.~\ref{fig:main}, the picture changes drastically. Vertically aligned dimers are no longer vacua of the theory, the domain
walls $2$ and $3$ and separately $1$ and $2$ of Fig.~\ref{fig:main} merge, forming a $\pi$ domain walls between the horizontal vacua offset
by a $Z_2$ shift. An isolated spin is then stuck on a domain wall interpolating between these two vacua, as in Fig.~\ref{fig:deconf}(d).
Such a deformation leads to a $\cos(2\chi)$ term exactly as in Eq.~\eqref{eq:YMeff} and the parallel with ``liberation on the wall''
in gauge theories \cite{Anber:2015kea} is complete.

\begin{figure}[tp]
\centering
\includegraphics[width=0.9\columnwidth]{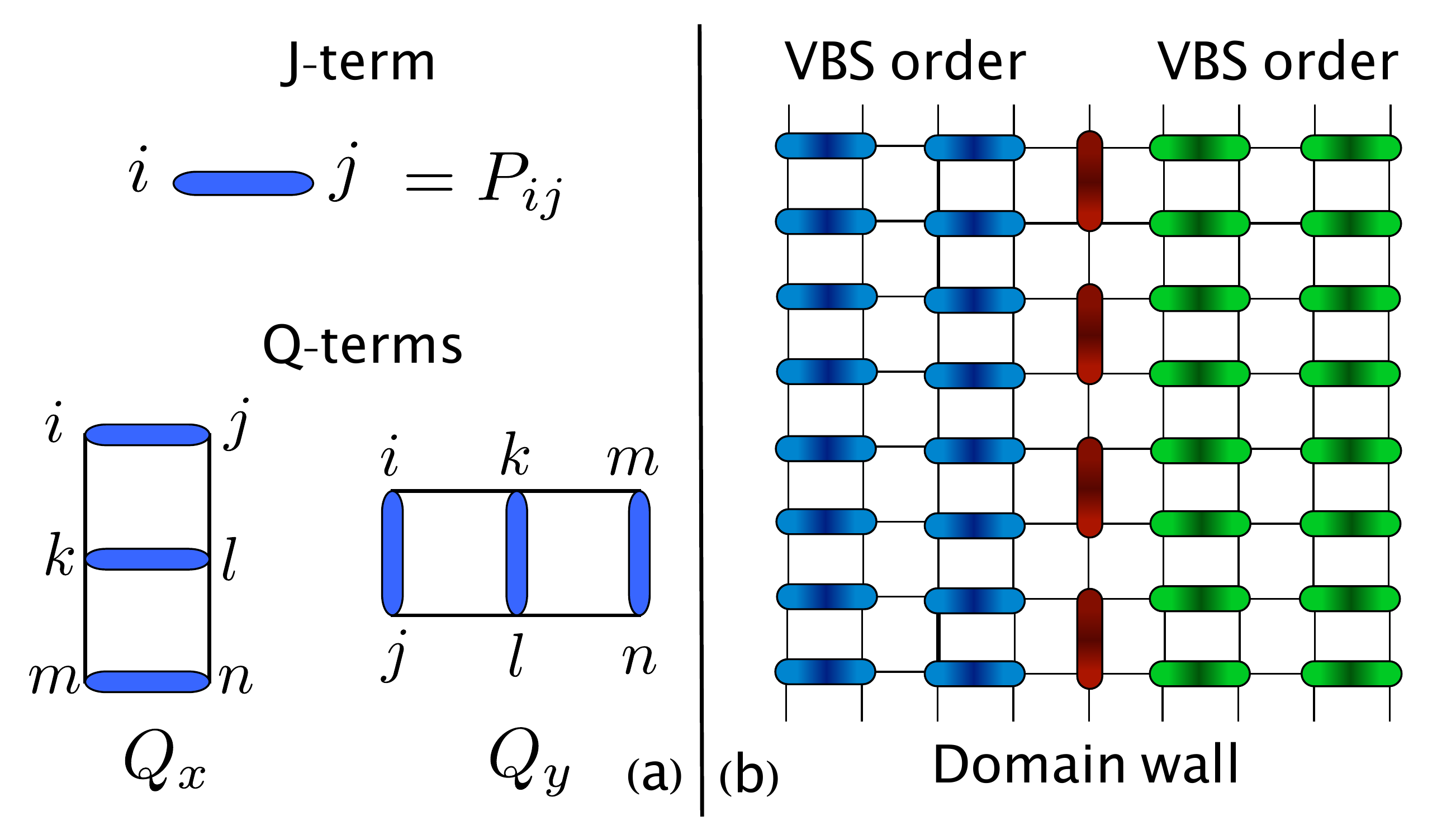}
\caption{The $J$-$Q$ model and domain walls. (a) The singlet projectors $P_{ij}$ of the $J$ and $Q$ terms. (b) Using periodic boundaries in the
$y$-direction and open boundaries in the $x$-direction, a $\pi$ domain wall is enforced when $L_x$ is odd and $Q_x>Q_y$. The $y$-direction is compactified.}
\label{fig:JQ}
\end{figure}

To numerically study a VBS domain wall we use the $J$-$Q$ model \cite{sandvik07}. Distinguished by the absence of QMC sign problem,
it has been used extensively \cite{kaul2013} to explore VBS states and deconfined criticality
\cite{melko08,lou09,sandvik12,harada13,block13,shao16}. The Hamiltonian $H=-JH_J-Q_xH_x -Q_yH_y$ contains singlet projectors
$P_{ij}=1/4-\bm S_i\cdot\bm S_j$ as explained in Fig.~\ref{fig:JQ}(a). The different $Q$-interactions for $x$- and $y$-oriented singlet projectors
allow us to study both $Z_4$ (for $Q_x=Q_y$) and $Z_2$ ($Q_x \not = Q_y$) VBSs. We use an unbiased ground-state QMC method
\cite{sandvik10,sandvik12,shao15,shao16} and set $Q_x=1$.

When $Q_x=Q_y=Q$, a deconfined transition takes place at $q=Q/(J+Q)\approx 0.6$; for $q>0.6$ the ground state is a $Z_4$ VBS on a torus of
size $L \times L$ with $L$ even \cite{lou09}. By setting $Q_y<Q_x$, open boundaries in the $x$ direction, and an odd length $L_x$,
the energetics lead to domain wall along the $y$-direction, as illustrated in Fig.~\ref{fig:JQ}(b). The domain wall is broadened by fluctuations
and is not fixed at the center of the system. The width of the domain wall is roughly the bulk dimer correlation length, which is less than the
lattice spacing deep inside the VBS phase ($J\ll Q_x,Q_y$) and grows as $J$ is increased. When $Q_y \not=Q_x$ the transition to the N\'eel
state is first-order \cite{block13}; here we are only concerned with the VBS.

\begin{figure}[tp]
   \centering
   \includegraphics[width=.9\columnwidth]{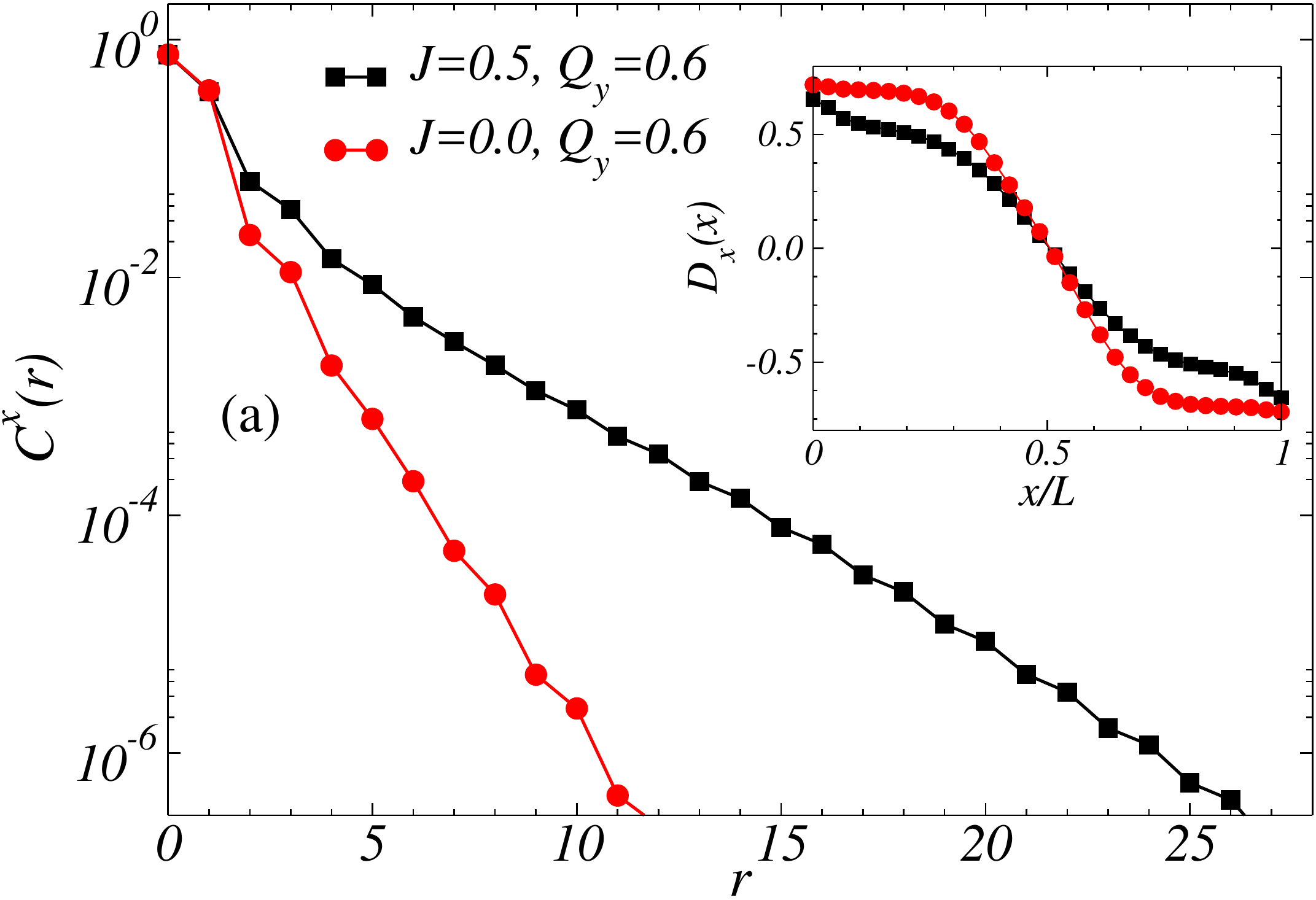}\\
   \vspace{0.3cm}
   \includegraphics[width=.9\columnwidth]{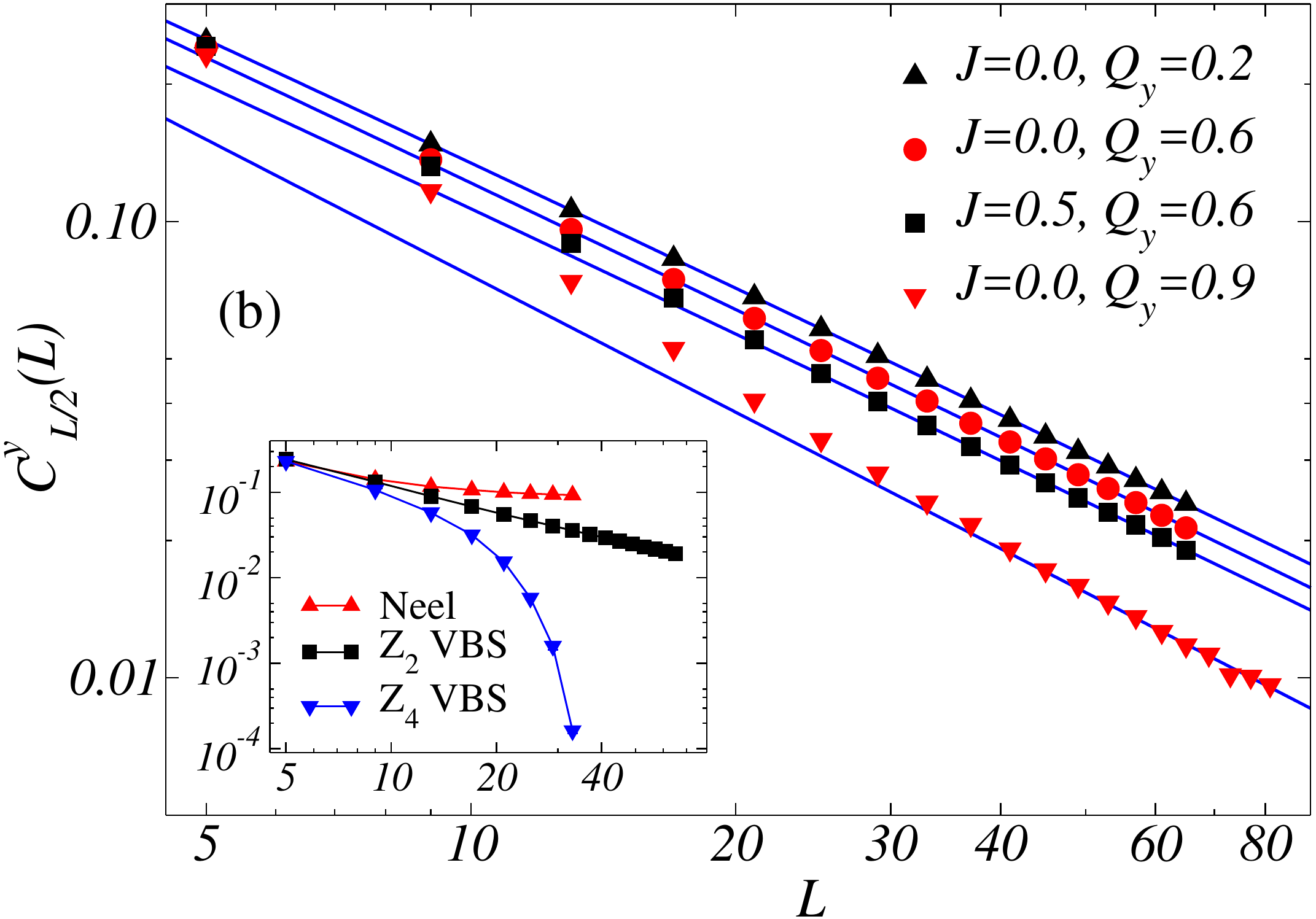}
   \caption{Correlations in the presence of a domain wall. The $x$ boundaries of the $(L+1)\times L$ $J$-$Q$ lattice ($L$ even)
     are open, which forces a domain wall in the $y$-direction. The coupling $Q_x=1$. (a) Spin correlations transverse
     to the domain wall at $Q_y=0.6$ and $L=32$. Averaging has been performed over all spin pairs separated by $(\Delta x=r,\Delta y=0)$.
     The inset shows the VBS (dimer) order parameter vs the lateral system coordinate. (b) Correlations parallel to the domain wall at
     $r=L/2$ fitted to the critical Heisenberg form. The inset shows the behavior in three different phases of the
     model: N\'eel-ordered ($J=5.0$, $Q_y=0.6$), $Z_2$ VBS ($J=0.5$, $Q_y=0.6$), and $Z_4$ VBS ($J=0.0$, $Q_y=1.0 $).}
\label{fig:correlators}
\end{figure}

It is possible to study a spinons using QMC in a basis of valence bonds and unpaired spins \cite{tang13}. In the present case, it is easier
to just confirm that the domain wall hosts a critical mode. With the domain wall along the $y$-direction, we expect the spin correlations in
the $x$-direction to decay exponentially with distance. This is demonstrated in Fig.~\ref{fig:correlators}(a) for two different
sets of couplings; one case where the host is deep in the VBS ($J=0,Q_y/Q_x=0.6$) and one case where the fluctuations are significant
($J/Q_x=0.5,Q_y/Q_x=0.6$). The inset panel shows the VBS order parameter \cite{sandvik12}, demonstrating explicitly the phase change
due to the domain wall.

To study correlations along the domain wall we define $C^y(r)=\langle {\bf m}(y)\cdot {\bf m}(y+r)\rangle$, where ${\bf m}$ is total
spin on a lattice row. Fig.~\ref{fig:correlators}(b) shows that the dependence on $r=L_y/2$ fits the two-point function of the critical
Heisenberg chain \cite{singh89}, $C^y(L/2) \sim L^{-1}\ln^{1/2}(L/L_0)$, from which we can infer that spinon excitations, although
confined in the bulk, are liberated on the domain wall.

The inset of Fig.~\ref{fig:correlators}(b) demonstrates explicitly that deconfinement does not take place in the $Z_4$ VBS, where the
system has two $\pi/2$ domain walls (as in Fig.~\ref{fig:main}) \cite{shao15}; the spin correlations decay exponentially, indicating a gap
and confined spinons. In the main panel of Fig.~\ref{fig:correlators}(b), the data at $Q_y=0.9$ exhibits a cross-over behavior, where
$L \agt 40$ is required to observe the critical Heisenberg behavior. For smaller $L$ the $\pi$ domain wall is not fully established.

In 3D the physics of the domain wall should be even richer. The membrane-like domain wall may host N\'eel order, in which case its excitations
are spin waves. However, depending on the host quantum magnet, the domain wall could also be a spin liquid with deconfined spinons. In addition
to possible realizations in magnetic solids, a natural setting to study domain-wall deconfinement experimentally with high tunability
would be optical lattices, where there are efforts underway to design quantum spin Hamiltonians \cite{greif13}. On a more fundamental level,
studies of various other aspects of confinement in quantum magnets, e.g., the nature of the confining string and its breaking when matter
is created (here spinons, but more generally fermions can be introduced by doping), may provide valuable information relevant also in QCD.

Realistic QCD regimes do not have degenerate vacua, but non-degenerate, so-called $k$-vacua most likely exist \cite{kvacua}. It is precisely
two or more of these vacua that become degenerate in the SUSY limit, or when the topological angle is dialed to $\theta=\pi$ (as we did here). Part of the string tension should be due to the excitation of these $k$-vacua, even in the regime which is
inaccessible to reliable computations. This would in turn imply that the topological charge density fluctuation is sensitive to the presence
of the QCD string, which can be tested on the lattice. The QCD string may have nontrivial interactions with the axion---a hypothetical particle
which, among other intriguing features, is a candidate for dark matter. An intriguing question is to what extent some of these open issues can
also be studied in quantum magnets, or in the richer setting of doped quantum magnets.

\begin{acknowledgments}
{\bf Acknowledgments:}
This research was supported by the DOE under Grants No.~DE-FG02-03ER41260 (TS) and DE-SC0013036 (M\"U) and by the NSF under Grant No.~DMR-1410126 (AWS).
HS would like to thank Boston University's Condensed Matter Theory Visitors program for support. TS and AWS would like to thank the European Centre for
Theoretical Studies in Nuclear Physics and Related Areas for support during the workshop \emph{Recent Advances in Monte Carlo Methods 2015}, where
the study was conceived. Some of the computations were carried out using Boston University's Shared Computing Cluster.
\end{acknowledgments}

\end{document}